\documentclass[a4paper]{article}
\usepackage{epsfig}
\usepackage{amsmath}
\usepackage{a4wide}
\numberwithin{equation}{section}
\newcommand{\half}{\frac{1}{2}}

\newcount\xrefpos \xrefpos=0
\newcount\yrefpos \yrefpos=0
\newcount\xput \xput=0
\newcount\yput \yput=0
\def\refpos#1 #2 #3{\global\xrefpos=#1 \global\yrefpos=#2
                         \rlap{$\smash{#3}$}}
\def\put #1 #2 #3{\xput=#1 \yput=#2
                  \advance\xput by -\xrefpos
                  \advance\yput by -\yrefpos
                  \rlap{\kern\the\xput truebp
                        \vbox to 0pt{\vss\hbox{$\displaystyle #3$}%
                        \kern\the\yput truebp}}}
\def\beginlabels\refpos#1\endlabels{\hbox{$\refpos#1$}}

\begin{document}
\title{The Gowdy $T^3$ Cosmologies revisited}
\thispagestyle{empty}
\author{S.D.~Hern and J.M.~Stewart \\
Department of Applied Mathematics \& Theoretical Physics, \\
Silver Street, Cambridge CB3 9EW, UK}

\date{30 January 1998}
\maketitle
\begin{abstract}
We have examined, repeated and extended earlier numerical calculations
of Berger and Moncrief for the evolution of unpolarized Gowdy~$T^3$
cosmological models.
Our results are consistent with theirs and we support their claim
that the models exhibit AVTD behaviour, even though spatial 
derivatives cannot be neglected.
The behaviour of the curvature invariants and the formation of 
structure through evolution both backwards and forwards in time is
discussed.



\medskip
\noindent DAMTP R-97/41

\noindent gr-qc/9708038

\noindent To appear in Classical and Quantum Gravity
\end{abstract}


\section{Introduction}
\label{sec:1}

The Hawking-Penrose theorems \cite{Hawking-Ellis} imply
that every `physically reasonable' cosmological model possesses a
singularity; however very little is known about its nature.
Early studies due to Belinskii \emph{et al}~\cite{BKL}
suggested that the behaviour of the generic singularity
resembles that of the spatially homogeneous Bianchi VIII and IX 
cosmological models, commonly called \emph{mixmaster dynamics}.
It is generally felt, however, that matters are not this simple, and
attention has focused on the \emph{velocity-term-dominated 
behaviour} (VTDB) hypothesis due to Eardley \emph{et al}~\cite{ELS}.
It is very hard to give a covariant definition but, loosely speaking,
the hypothesis claims that near the singularity spatial derivatives
play no part in the evolution; locally one has a spatially homogeneous
cosmology, but with the parameters varying from point to point.
This is the starting point for many large-scale-structure calculations
in physical cosmology (see, for example,~\cite{SS}), but is it
correct? 
If spatial derivatives are ignored, the evolution equations become
ordinary differential equations whose behaviour is easier to analyse.
Isenberg and Moncrief~\cite{IM} have introduced an alternative
hypothesis of \emph{asymptotically velocity-term dominated} (AVTD)
behaviour.
A cosmological model is AVTD if its asymptotic behaviour near the
singularity is that of the ordinary differential equation system.
Clearly VTDB implies AVTD behaviour, but the converse may not be true.
In order to test the validity and usefulness of these hypotheses we 
need to examine the behaviour of inhomogeneous cosmological
models, and the simplest examples would appear to be the 
Gowdy~$T^3$ models~\cite{Gowdy}.
Polarized Gowdy cosmologies exhibit AVTD behaviour~\cite{IM},
and so attention focuses on the less tractable unpolarized
case~\cite{GM}, which appears to require numerical treatment.

Berger and Moncrief~\cite{BM} and Berger~\cite{B97} (who give
much more extensive bibliographies than the brief introduction above)
have tackled the unpolarized case numerically. 
This is a decidedly non-trivial task, and so they introduced novel 
numerical algorithms, which led them to conclude, tentatively, that 
the AVTD hypothesis was confirmed.
Their grounds for caution are based on the inability of their code to
resolve fine-scale spatial structure.
As they suggest, this problem requires an algorithm with adaptive mesh
refinement (AMR), discussed in section 3.

We have used our AMR code to repeat and extend the earlier calculations.
We obtain broadly similar results, but with additional fine structure.
We confirm the earlier result that the Gowdy~$T^3$ cosmologies exhibit
AVTD behaviour but can find no evidence that the VTDB hypothesis holds
universally near the singularity.
Our results show that by starting with smooth data and evolving backwards
in cosmic time towards the singularity our fields develop complicated
spatial structure.
If we time-reverse our calculations we appear to be 
asserting that complicated
structure becomes smooth as it evolves into the future, the exact
opposite of widespread beliefs in the theory of large-scale structure!
We have therefore taken generic smooth data near the singularity
and evolved it into the future finding, as expected, that complicated
spatial structure appears.
The fields that we are discussing are merely metric components with
no covariant meaning.
We have therefore computed curvature invariants.
These become very large near the singularity, but they too have
complicated spatial structure, implying that this is not a coordinate
effect.

\section{The Gowdy $T^3$ Universe}
\label{sec:2}

The line element used by Berger and  Moncrief \cite{BM} for the Gowdy $T^3$
cosmology is
\begin{equation}
  \label{eq:line}
  \begin{split}
    ds^2 = {}
    &e^{\lambda/2}e^{\tau/2}(-e^{-2\tau}d\tau^2 + d\theta^2) +\\
    &e^{-\tau}(e^P\,d\sigma^2 + 2e^P Q \,d\sigma\,d\delta +
    (e^P Q^2 + e^{-P})\,d\delta^2),
  \end{split}
\end{equation}
where $\lambda$, $P$ and $Q$ are functions of $\tau$ and $\theta$ only, 
$-\infty < \tau < \infty$ with $\tau = \infty$ a singularity,
and $0 \leq \theta, \sigma, \delta \leq 2\pi$.
The functions $\lambda$, $P$ and $Q$ are required to be periodic in
$\theta$ with period $2\pi$.
The polarized mode corresponds to $Q=0$.

The vacuum momentum constraint equation is 
\begin{equation}
  \label{eq:con1}
  - \lambda_\theta = 2(P_\tau P_\theta + e^{2P}Q_\tau Q_\theta),
\end{equation}
and the Hamiltonian constraint is
\begin{equation}
  \label{eq:con2}
  - \lambda_\tau = P_\tau{}^2 + e^{-2\tau}P_\theta{}^2 +
  e^{2P}(Q_\tau{}^2 + e^{-2\tau}Q_\theta{}^2),
\end{equation}
while the vacuum evolution equations reduce to
\begin{subequations}
  \label{eq:evol}
  \begin{align}
    P_{\tau\tau} &= e^{-2\tau}P_{\theta\theta} +
    e^{2P}(Q_\tau{}^2 - e^{-2\tau}Q_\theta{}^2),
    \\
    Q_{\tau\tau} &= e^{-2\tau}Q_{\theta\theta} -
    2(P_\tau Q_\tau - e^{-2\tau}P_\theta Q_\theta).
  \end{align}
\end{subequations}
Here $f_\tau = \partial f/\partial \tau$, etc.
It should be noted that the evolution equations~\eqref{eq:evol}
do not involve $\lambda$. 
The energy constraint \eqref{eq:con2} determines $\lambda$ and the
momentum constraint \eqref{eq:con1} is satisfied at all times if it is
satisfied initially, provided the other equations hold.
As pointed out by Berger and Moncrief~\cite{BM} the evolution 
equations are harmonic map equations for a target space with metric
\begin{equation}
  \label{eq:harm}
  dS^2 = dP^2 + e^{2P}\,dQ^2,
\end{equation}
and so study of these equations is not without interest.

The integration programmes described in the next section require a
first order system of equations.
We accomplish this by introducing new variables as follows:
\begin{equation}
  \label{eq:vars}
  \quad A = P_\tau, \quad B = Q_\tau, \quad
  C = P_\theta, \quad D = Q_\theta.
\end{equation}
The system of evolution equations now takes the form
\begin{subequations}
  \label{eq:first}
  \begin{align}
    A_\tau &= e^{-2\tau}C_\theta + e^{2P}(B^2 - e^{-2\tau}D^2),\\
    B_\tau &= e^{-2\tau}D_\theta - 2(AB - e^{-2\tau}CD),\\
    C_\tau &= A_\theta,\\
    D_\tau &= B_\theta,\\
    P_\tau &= A,\\
    Q_\tau &= B,\\
    \lambda_\tau &= - A^2 - e^{-2\tau}C^2 
                    - e^{2P}( B^2 + e^{-2\tau}D^2 ).
  \end{align}
\end{subequations}
The last equation is only needed if~$\lambda$ is to be evolved
simultaneously with~$P$ and~$Q$, in which case
\begin{equation}
  \label{eq:constr}
  \lambda_\theta = -2(AC + e^{2P}BD)
\end{equation}
from equation~\eqref{eq:con1} acts as a constraint.

Initial data at $\tau=0$ is also required.  
Berger and Moncrief suggest that the following choice is reasonably
generic (with~$\lambda$ chosen to satisfy~\eqref{eq:constr}):
\begin{equation}
  \label{eq:cauchy}
  A = v_0\cos\theta, \quad B = 0, \quad C = 0, \quad 
  D = -\sin\theta, \quad P = 0, \quad Q = \cos\theta, \quad
  \lambda = 0,
\end{equation}
and the results reported here use $v_0 = 10$ as used in \cite{BM}.

\section{Numerical Procedures}
\label{sec:3}

As we shall see in the next section, the evolution produces structure
on fine scales with steep gradients.
One numerical strategy is to use a standard algorithm, e.g.,
Lax-Wendroff, together with a very fine grid structure.
This certainly works  but it is not the most efficient way to proceed.
Berger and Moncrief~\cite{BM} used a \emph{symplectic integrator},
described in their paper.
However their calculation produced fine-scale structure even at sizes
comparable with the grid spacing, which they regarded as unreliable.
In presenting their results spatial averaging was used to remove the
finest-scale structure.
An alternative approach is due to van Putten \cite{Putt}, but he 
reports integration of the equations only for short times and small
values of the parameter $v_0$ in the initial data.

Problems of this type are natural candidates for 
\emph{adaptive mesh refinement} (AMR), 
which seeks to add extra grid points when they are
needed and to remove them when they become unnecessary.  
We used an implementation of the Berger and Oliger~\cite{BO} AMR algorithm
which is descended from the one described in
Hamad\'e and Stewart~\cite{HS}.
Numerous changes have been made to that code.
Those relevant to this paper are:
\begin{itemize}
\item an arbitrary number of grids may exist at each level of
refinement;
\item it is now possible to switch in arbitrary integrators, giving the
code `modularity';
\item cubic spline interpolation is preferred to the quadratic
  interpolation used earlier, thus guaranteeing continuous spatial
  second derivatives;
\item when creating finer grids one can choose to use data
  from obsolete grids at the same level, rather than relying on
  interpolation from the parent grid.
\end{itemize}
For non-specialists, the third change introduces a subtle form of
data smoothing, which can be (almost) eliminated by the fourth change.

It is straightforward to write a generalization of the standard
Lax-Wendroff two-step algorithm which is second-order accurate for
the system \eqref{eq:first}, and this proved to be the fastest choice.
(If the typical spatial grid spacing is $\Delta \theta$, then a 
second-order accurate method produces a local truncation error which is 
$O((\Delta \theta)^3)$.
The mesh refinement is typically 
$\Delta \theta \rightarrow \Delta \theta/4$,
and so a single stage of refinement reduces the local truncation error
by nearly two orders of magnitude, as well as enhancing spatial 
resolution.)
In addition (and as part of a long-term programme) we used our
interpretation of the second-order accurate wave propagation routine 
from the \textsc{clawpack} high-resolution package of LeVeque \cite{RJL}.
(\textsc{clawpack} is written in \textsc{fortran}.
The administrative part of our code is written in C++, and the
numerically intensive functions are written  either in C or C++. 
It proved easier, and more instructive, to adapt the \textsc{clawpack}
routines to this environment, rather than to interface the
\textsc{fortran} to the existing code.)

For calculations using the Berger-Oliger algorithm the initial coarse
grid had a spatial separation $\Delta \theta = 2\pi/2000$.
The finest child grid actually used had $\Delta \theta = 2\pi/512\,000$.
As a check to ensure that the AMR code was not introducing spurious
effects we repeated the calculations with $\Delta \theta = 2\pi/8000$ and
no mesh refinement.
The Lax-Wendroff and \textsc{clawpack}-like codes were developed by
the two different authors, continuing our policy to eliminate
coding errors. 
We also checked our codes against the exact pseudo-unpolarized solution 
\begin{equation}
  \label{eq:exact}
  P = \log \cosh P_0, \qquad Q = \tanh P_0,\quad\quad 
P_0(\tau, \theta) = J_0(e^{-\tau})\cos\theta,
\end{equation}
(where $J_0$ is a Bessel function) given by 
Berger and Moncrief~\cite{BM}, and confirmed that for both of our
integration methods the errors decreased at second order.
There were only very small differences between the results
from AMR with Lax-Wendroff and those from AMR
with \textsc{clawpack}, and what is reported in the next section
is common to both.

\section{Numerical Results}
\label{sec:4}

We are reporting on the integration of equations~\eqref{eq:first} with
initial data~\eqref{eq:cauchy} and $v_0 = 10$, the same conditions 
as used by Berger and Moncrief~\cite{BM}.
(More or less spatial structure develops for larger or smaller,
respectively, values of the parameter $v_0$.  
Berger~\cite{B97} uses $v_0 = 5$ and gets results with a less 
complicated spatial structure, though with the same basic features.)

\begin{figure}[tb]
  \begin{center}
    \vskip 3mm
    \vbox {
      \beginlabels
      \refpos 90 643 {}
      \put 96 601 {P}
      \put 96 516 {Q}
      \put 96 432 {\lambda}
      \put 208 378 {\theta}
      \put 408 378 {\theta}
      \put 198 643 {\tau = 6 \pi}
      \put 393 643 {\tau = 54 \pi}
      \endlabels
      \includegraphics[width=14cm,height=9cm]{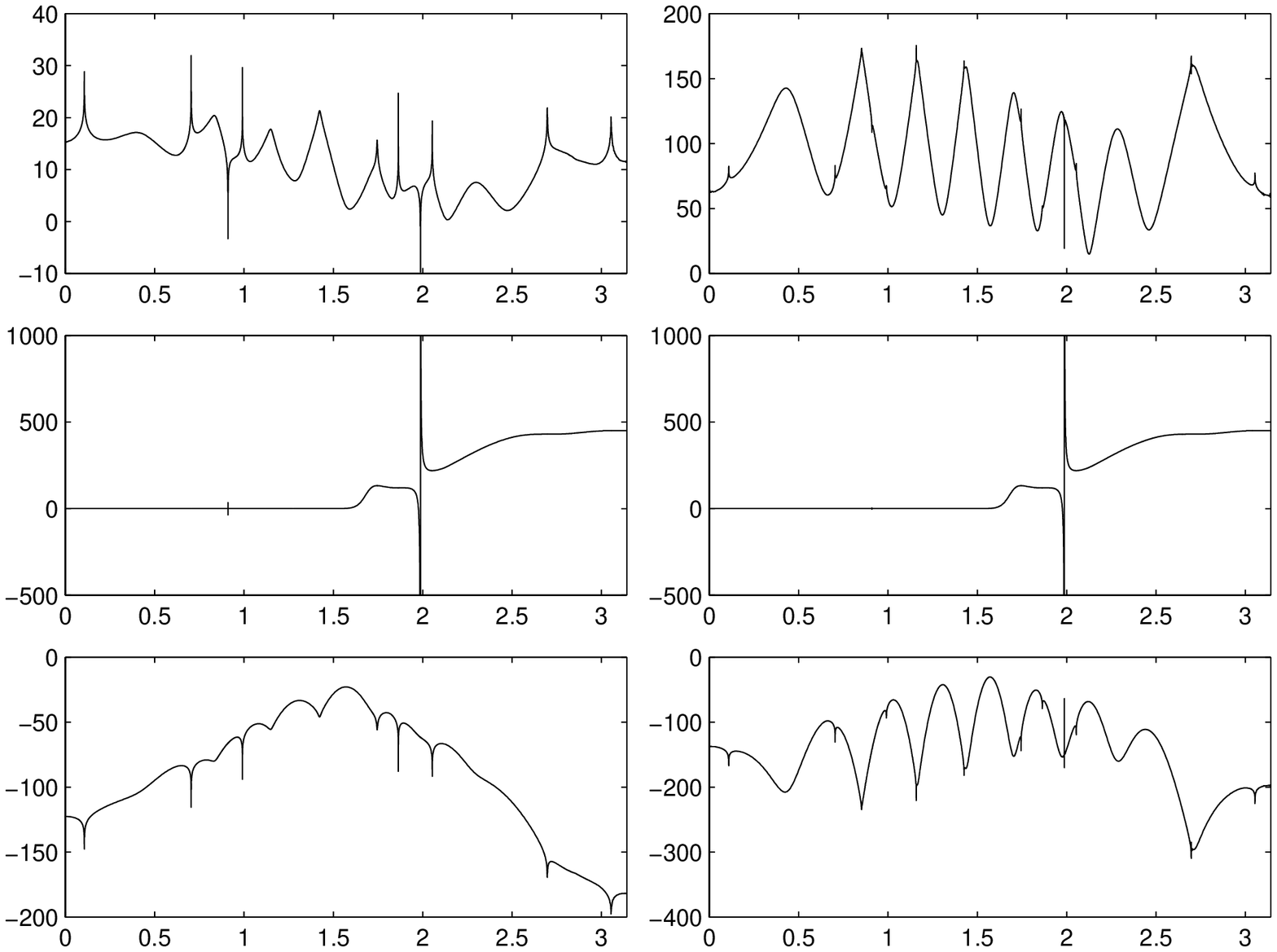}
    }
    \vskip 5mm
    \begin{quote}
      \small Figure 1:
      $P$, $Q$ and $\lambda$ as functions of $\theta$
      at two different times.
      The left column corresponds to $\tau = 6\pi$,
      the right to $\tau = 54\pi$,
      and $\theta$ ranges from $0$ to $\pi$. 
      (The problem is invariant under $\theta \rightarrow 
      2\pi - \theta$, and so only one half of the full
      $\theta$-range is shown.)
      The large spikes in~$Q$ are truncated by the choice
      of vertical scale.
    \end{quote}
  \end{center}
\end{figure}

The main numerical results of Berger and Moncrief are summarized 
in figures~4 and~5 of~\cite{BM}.
The left-hand column of our figure~1 shows $P$, $Q$ and $\lambda$
as functions of $\theta$ 
at $\tau = 6 \pi = 18.85$, the latest time of their figures~4 and~5.
The right-hand column of figure~1 shows the variables at a much later 
time of $\tau = 54 \pi = 169.6$.
The graphs in figure~1 comprise all available data
from the AMR grid hierarchy. 

\bigskip

The structure of $P$ becomes complicated soon after
the start of the evolution
and by time $\tau = 6 \pi$ (as figure~1 shows) it includes a number of
fine spikes.
At first sight these spikes might seem to be errors in the
evolution, and, indeed, when Berger and Moncrief first found spikes
in their data, they considered them unreliable and used spatial 
averaging to remove them from figures~4 and~5 of~\cite{BM}.
However, our AMR code automatically refines the evolution in
regions where spikes start to form, increasing the grid
resolution so that the spikes appear smooth and the evolution
can continue without loss of accuracy. 
This is demonstrated in figure~2 where data at a spike is
shown for a coarse grid and for two refined subgrids.
(Notice that on the finer grids the spike reaches a greater height
than on the coarse grid. The AMR 
algorithm uses an average of data from fine grids to update 
data on coarse grids.) 

\begin{figure}[tb]
  \begin{center}
     \vskip 5mm
     \vbox {
       \beginlabels
       \refpos 90 725 {}
       \put 91 669 {P}
       \put 158 727 {\hbox{Parent}}
       \put 292 727 {\hbox{Child}}
       \put 410 727 {\hbox{Grandchild}}
       \put 170 600 {\theta}
       \put 301 600 {\theta}
       \put 432 600 {\theta}
       \endlabels
       \includegraphics[width=14cm,height=4cm]{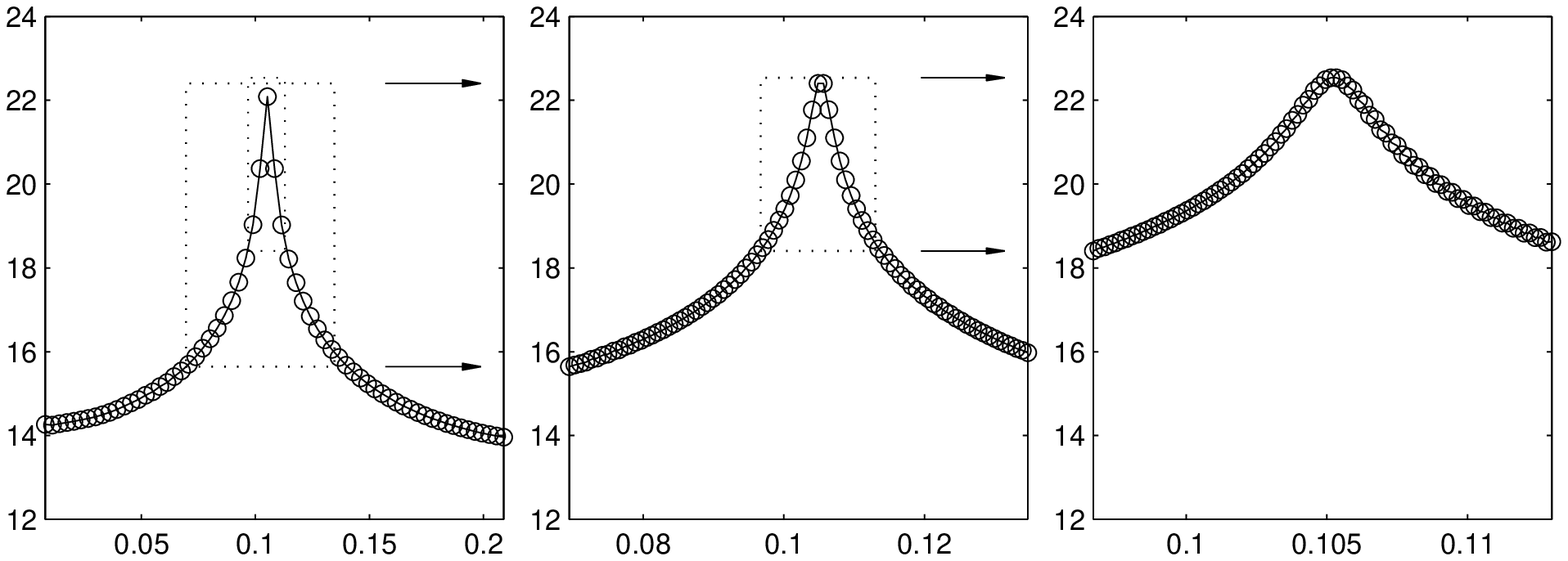}
     }
     \vskip 5mm
     \begin{quote}
       \small Figure 2:
       Data from three generations of grids at a spike in~$P$.
       Each grid has a resolution greater by a factor of four than 
       the grid to its left.
       Each circle is one grid point, and the dotted boxes show
       the sizes and data ranges of the subgrids.
       (The data is taken at time $\tau = 15$.)
     \end{quote}
  \end{center}
\end{figure}

Spikes form in $P$ in both the positive and negative directions.
Examining the development of the positive spikes (the
negative spikes are discussed below) it is found that their peaks
grow linearly at rates $P_\tau = \hbox{constant} > 1$, while at the
same time they decrease in width.
As a spike narrows (and there is no indication that the spikes ever
stop growing or narrowing) the AMR code is forced to use finer and
finer grids to keep it well resolved until, eventually, 
it becomes narrower than the smallest grid spacing
that the AMR code has been instructed to use, 
and from that point on it cannot be assumed that the code is 
accurately evolving the spike.
We have to accept that there are probably significant errors in
the calculated heights of the spikes in $P$, but this does not imply
that there must be large errors elsewhere in the data: since the
characteristic speed of the model is $\exp(-\tau)$, 
deviations from the exact solution that develop at some time not too
soon after the start of the
evolution can only propagate very small
distances from their points of origin. 

At late times $P_{\tau\tau} \equiv A_\tau$ becomes very small,
and so~$A$ freezes and~$P$ grows linearly in time with the spatial 
profile shown in the right-hand column of figure~1.
Except at the points where spikes formed, the spatial profile of~$P$
at late times has quite a simple form, much less intricate
than early on in its evolution.
For reasons discussed above, the numerical simulation is
unable to determine the late-time behaviour of the spikes.
While small spikes and lumps appear in the graph of~$P$ at
$\tau = 54 \pi$ in figure~1, these are just the residue left by
spikes that have narrowed beyond the resolution of the simulation.
The spikes themselves, assuming they continued to grow at their
initial rates, would go off the scale of the graph.

To suggest reasons for the observed behaviour it is
useful to know which are the dominant terms in the governing
equations.
Here we adopt the definition that a term is \emph{dominant} if its 
absolute value exceeds ten times the sum of the absolute values 
of the other terms.
Figure~3 illustrates this for the wave 
equations~(\ref{eq:first}a,b) for
$A_\tau \equiv P_{\tau\tau}$ and $B_\tau \equiv Q_{\tau\tau}$. 
The lightly shaded regions in $(\tau, \theta)$-space are where the
terms $\exp(2P)B^2$ (in (\ref{eq:first}a)) 
and $-2AB$ (in (\ref{eq:first}b)) dominate the
right-hand sides.
The darker shaded regions are where the terms $-\exp(2P-2\tau)D^2$ and
$2\exp(-2\tau)CD$ dominate.
The narrow black regions are where the terms
$\exp(-2\tau)C_\theta$ and $\exp(-2\tau)D_\theta$ dominate.
The white regions are where no one term dominates.
It is clear from the figure that there is very complicated interplay
between terms during the early stages of the evolution, and it seems
unlikely that the overall behaviour can be traced to simple causes.
Berger~\cite{B97} notes that spikes form at points where
$D \equiv Q_\theta = 0$ (though not all such zeros produce spikes)
and discusses their formation in terms of bounces off of potentials
in the evolution equations.
In terms of dominance behaviour, spike formation occurs in dark-grey
regions in the left half, and light-grey regions in the right half of
figure~3.

\begin{figure}[tb]
  \begin{center}
     \vskip 3mm
     \vbox {
       \beginlabels
       \refpos 90 725 {}
       \put 224 725 {A_\tau}      
       \put 379 725 {B_\tau}      
       \put 137 605 {\tau}      
       \put 225 476 {\theta}
       \put 380 476 {\theta}
       \endlabels
       \includegraphics[width=11cm]{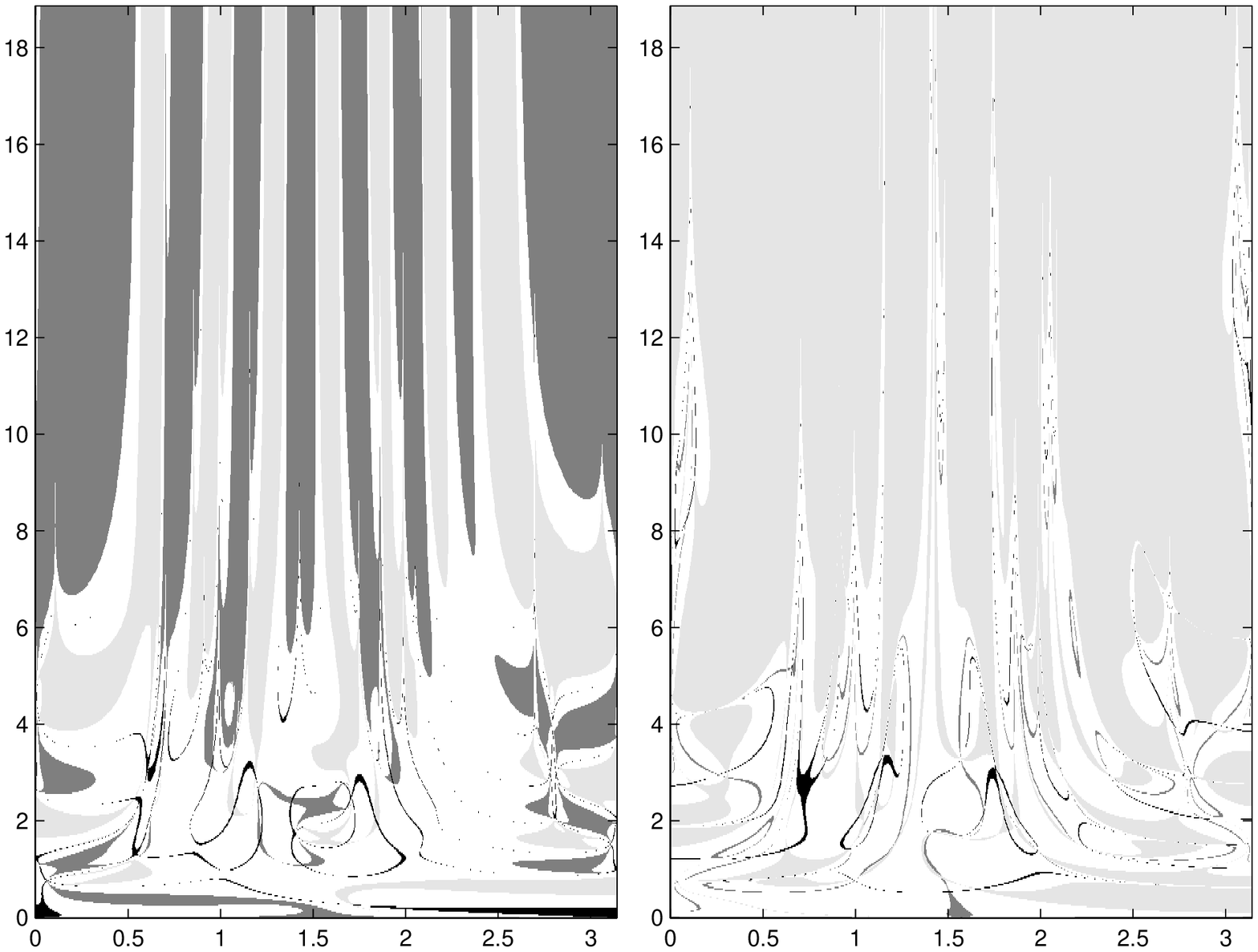}
     }
     \vskip 5mm
     \begin{quote}
       \small Figure 3:
       The dominant terms in the $A_\tau$ and $B_\tau$
       equations (\ref{eq:first}a,b) as functions
       of~$\theta$ and~$\tau$.
       The black regions are where terms involving $C_\theta$ 
       and $D_\theta$ dominate,
       the dark-shaded regions are where terms involving $C$ and $D$
       dominate, 
       and the light-shaded regions are where terms 
       involving $A$ and $B$ dominate.
       In the white regions no single term dominates.
     \end{quote}
  \end{center}
\end{figure}

\bigskip

Sharp features also form in~$Q$ as it evolves.
Figure~1 shows that~$Q$ at $\tau = 6 \pi$ has a negative spike
and a positive spike very close together with a
steep gradient between the two peaks.
A second, similar double-spike feature is also present, but
only just visible in figure~1 because of the scale.
The positions of the two double spikes in~$Q$ coincide with the
positions of the two negative spikes in~$P$.
A close examination of the data reveals that both the positive
and negative peaks of each double spike in~$Q$ grow at an
exponential rate, while the overall width of the feature
becomes smaller,
and that the negative spikes in~$P$ grow linearly
in much the same way that their positive counterparts do.
The discussion above regarding the numerical problems caused by 
the narrowing of spikes applies equally here.

A double-spike feature is also reported in~\cite{B97} (see 
especially figure~2 there, but note that it is produced 
using different initial data),
however the explanation given there is not quite convincing.
Examination of the dominance data of figure~3,
together with similar data for finer grids,
reveals that the region around a double spike
($\theta \approx 2$, $\tau > 4$ for the large double spike
in figure~1) is light grey for both equations.
We may conclude that in this region the terms proportional
to $\exp(-2\tau)$ are irrelevant and so
\begin{equation}
  \label{eq:approx}
  A_\tau \approx e^{2P}B^2, \qquad B_\tau \approx -2AB.
\end{equation}
A double spike then forms under the conditions of~$B$ crossing
zero (the exact position of this shows up as a thin white
line in both halves of figure~3), while~$A$ is negative
and~$P$ is negative or small and positive.
It follows that $A_\tau \geq 0$ and is small, so that~$A$
remains negative and~$P$ forms a negative spike.
On both sides of the zero,~$B$ then grows
exponentially, but in opposite directions, and the result
is a double spike in~$Q$.
Since $A_\tau \geq 0$ the region in which $A < 0$ becomes
smaller, and the double spike narrows.
(By $\tau = 10$ the light-grey region that supports the growth
of the larger double spike is no longer visible in the left
half of figure~3.)

Away from the double spikes, $Q_\tau \equiv B$ quickly approaches
zero and~$Q$ freezes not long after the start of the evolution:
the two plots of~$Q$ in figure~1 differ only at the double spikes. 
Because of the narrowing effect, 
the second double spike is still barely visible on the right half
of figure~1 although, extrapolating its initial growth, it should
be larger than the vertical scale of the graph.

\bigskip

The evolution of~$\lambda$ shows similar features to the
evolution of~$P$ (see figure~1).
Negative spikes develop in~$\lambda$ at the same points at which
positive spikes develop in~$P$ and, apart from pointing
in a different direction, they show the same behaviour:
they grow at a constant rate and they narrow until they no
longer can be resolved by the simulation.
In contrast, no unusual behaviour seems to develop in~$\lambda$
at the points where double spikes form in~$Q$ (although the
errors in the data at the double spikes are sufficient to
produce erratic behaviour in~$\lambda$ there
at late times, as can be seen in figure~1).

Equation (\ref{eq:first}g) shows that~$\lambda$ is a
decreasing function of time, and it is clear from figure~1
that~$\lambda$ becomes large and negative as it evolves.
At late times, $\lambda_\tau \approx - A^2$ 
(cf.~equation~(\ref{eq:first}g)) and~$A$ is approximately constant,
so~$\lambda$ grows linearly with a fixed spatial profile.

The constraint equation~\eqref{eq:constr} in our system relates the 
spatial derivative of~$\lambda$ to the values of~$P$ and~$Q$ and 
their first derivatives.
For an exact solution of the evolution equations~\eqref{eq:first}
the constraint equation is satisfied at all times if it is 
satisfied initially, but for a numerical solution this is not
necessarily the case.
We monitor the degree to which the constraint equation fails to
hold by calculating the total percentage error defined by
\begin{equation}
  \label{eq:conserr}
  \hbox{error(\%)} = 100 
    \int | \lambda_\theta + 2(AC + e^{2P}BD)| \, d \theta
    \, \bigg/
    \int | \lambda_\theta | \, d \theta ,
\end{equation}
where the integrals are evaluated using the trapezium rule
and~$\lambda_\theta$ is estimated from~$\lambda$ using
second-order finite differencing.
Since the errors are known to be disproportionately large
at the spikes we do not include these points when 
measuring the constraint.
(A region of width~0.02 is excised at each spike, with
about ten percent of the total domain being removed.)
At time $\tau = 6 \pi$ the error in the constraint 
is~1.1\%. 
At time $\tau = 54 \pi$ the error is~0.97\%, smaller because
of the overall growth of~$\lambda$.

Instead of evolving~$\lambda$ with equation~(\ref{eq:first}g)
we could calculate it at every time step by integrating
equation~\eqref{eq:constr}.
This would obviously ensure that the constraint equation is
satisfied, and such `fully constrained' approaches are often
applied to problems in numerical relativity.
However, for this particular problem enforcing the constraint
equation in this way would be a very bad idea:
the large errors at the spikes would affect the whole
of~$\lambda$ rather than being confined to isolated regions.

\bigskip

\newcommand{\inva}{w_{\hbox{\scriptsize I}}}
\newcommand{\invb}{w_{\hbox{\scriptsize II}}}
\newcommand{\invc}{w_{\hbox{\scriptsize III}}}
\newcommand{\invd}{w_{\hbox{\scriptsize IV}}}

Our understanding of a spacetime is, in general, advanced more
by studying coordinate-independent quantities which have some
physical relevance than by studying the components of the metric.
For a vacuum spacetime we can use the curvature tensor to
construct four independent scalar quantities---the curvature
invariants---which we take in the form
\begin{equation}
  \label{eq:invars}
  \begin{array}{cc}
    \displaystyle
    \inva = \frac{1}{8} C_{\mu\nu\rho\sigma}
                      C^{\mu\nu\rho\sigma}
    , \quad &
    \displaystyle
    \invc = -\frac{1}{16} C_{\mu\nu}{}^{\rho\sigma}
                        C_{\rho\sigma}{}^{\eta\omega}
                        C_{\eta\omega}{}^{\mu\nu}
    , \\*[3ex]
    \displaystyle
    \invb = \frac{1}{8} C^{\textstyle\ast}_{\mu\nu\rho\sigma}
                      C^{\mu\nu\rho\sigma}
    , \quad &
    \displaystyle
    \invd = -\frac{1}{16} C^{\textstyle\ast}_{\mu\nu}{}^{\rho\sigma}
                        C_{\rho\sigma}{}^{\eta\omega}
                        C_{\eta\omega}{}^{\mu\nu}
    ,
  \end{array}
\end{equation}
where $C_{\mu\nu\rho\sigma}$ is the Weyl tensor,
$\varepsilon_{\mu\nu\rho\sigma}$ is the 
Levi-Civita tensor
and $C^{\textstyle\ast}_{\mu\nu\rho\sigma} \equiv 
\half \varepsilon_{\mu\nu\eta\omega} C^{\eta\omega}{}_{\rho\sigma}$.

Figure~4 shows two of the curvature invariants, $\inva$
and~$\invb$, evaluated using our data at time $\tau = 6 \pi$.
The following discussion concentrates on these two quantities
since~$\invc$ and~$\invd$ behave qualitatively the same as,
respectively,~$\inva$ and~$\invb$.

\begin{figure}[tb]
  \begin{center}
    \vskip 3mm
    \vbox {
      \beginlabels
      \refpos 90 725 {}
      \put 206 725 {\inva}      
      \put 409 725 {\invb}      
      \put 208 602 {\theta}
      \put 411 602 {\theta}
      \endlabels
      \includegraphics[width=14cm,height=4cm]{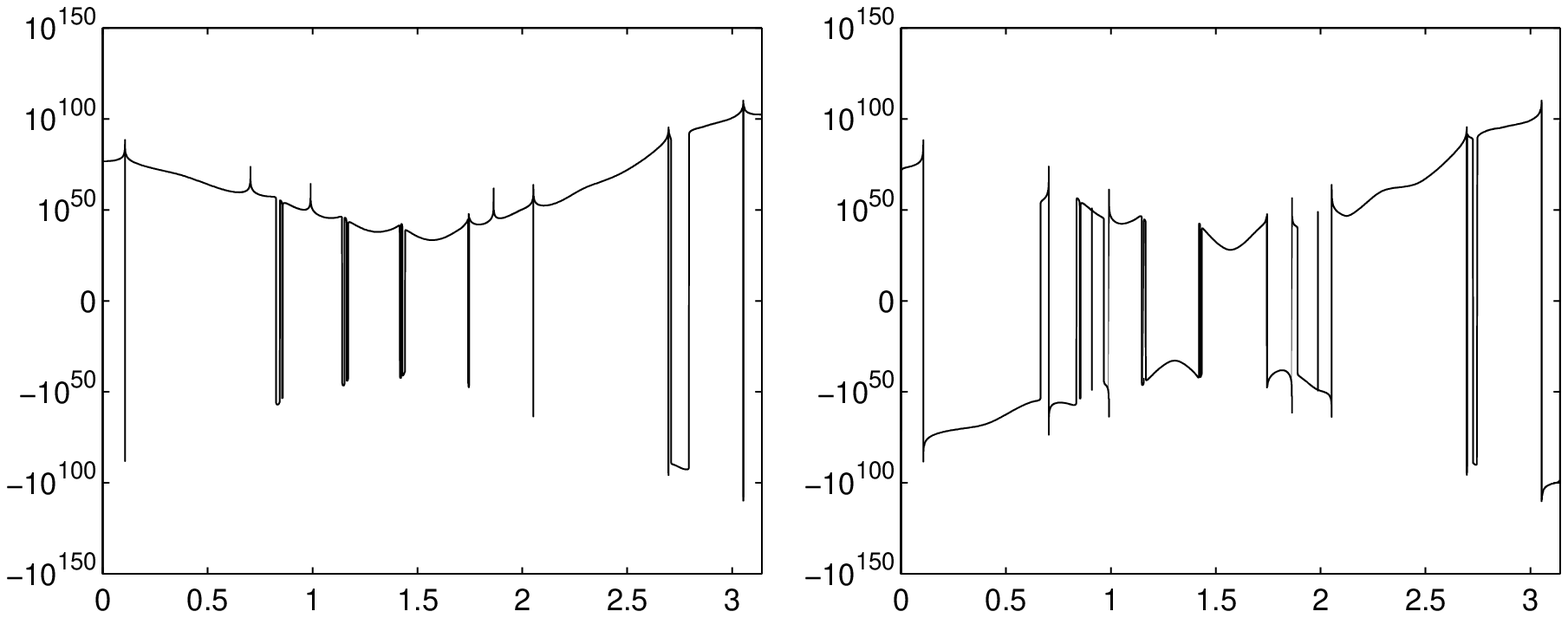}
    }
    \vskip 3mm
    \begin{quote}
      \small Figure 4:
      Curvature invariants~$\inva$ and~$\invb$
      (see equations~\eqref{eq:invars}) at $\tau = 6 \pi$
      for $0 \leq \theta \leq \pi$.
      While $\inva$ is invariant under the reflection
      $\theta \rightarrow 2\pi - \theta$, the sign
      of $\invb$ is reversed.
      (The data has been scaled using the transformation
      $w \rightarrow (1 / \ln 10) \sinh^{-1} (w/2)$.) 
    \end{quote}
  \end{center}
\end{figure}

The most striking feature of the curvature invariants is
their magnitude:~$\inva$ and~$\invb$ at $\tau = 6 \pi$
reach both positive and negative values of size greater than
ten to the hundredth power.
At the start of the evolution~$\inva$ takes values from $-50$
to $+30\,000$, and~$\invb$ from $-2500$ to $+2500$.
Their subsequent growth is rapid, and shows no signs of reversing.
The main influence on the size of the curvature invariants
is the factor $\exp(-\lambda)$ in the
expressions for~$\inva$ and~$\invb$.
The effect of~$\lambda$ can be seen by comparing the profiles of~$\inva$
in figure~4 and~$\lambda$ at $\tau = 6\pi$ in figure~1.
The linear growth (towards minus infinity) of~$\lambda$ 
at late times corresponds to an
exponential growth of the curvature invariants.

As figure~4 shows, the curvature invariants change sign suddenly
at various values of~$\theta$.
These sign changes tend to form into clusters which contract
in width as the evolution progresses.
In~$\inva$ the clusters consist of even numbers of sign changes
and the function is predominantly positive, while in~$\invb$ they are
odd numbered and the function changes sign between clusters.
The sign change clusters become narrow in just the same way that
the spikes in~$P$ do, and indeed the two types of feature 
coincide.
The narrowing, together with the inaccuracies in the data at
the spikes, makes determining the precise behaviour of sign 
changes within a cluster difficult.
However, careful examination suggests that while the width of a
cluster tends to zero, there is no cancellation between the sign
changes.

Given the large numbers of terms in the expressions for the
curvature invariants, relating their behaviour to features of
$P$,~$Q$ and~$\lambda$ is not a simple matter.
However, by comparing the magnitudes of individual terms we have found
that at late times we can approximate the curvature invariants with
simplified expressions
\begin{subequations}
  \label{eq:approxinv}
  \begin{align}
    \inva
    & \displaystyle
    \approx
    \frac{1}{32} e^{3\tau-\lambda} (A^2+3) (A^2-1)^2 , 
    \\
    \invb 
    & \displaystyle
    \approx
    - \frac{1}{16} e^{2\tau+P-\lambda} D A (A+3) (A-1) (A+1)^2 .
  \end{align}
\end{subequations}
These approximations fail when the values they predict are 
small: other terms in the curvature invariants clearly cannot be 
neglected then.
The sign changes in~$\inva$ are all clustered around the positive
spikes in~$P$, and this makes sense in terms of the approximation
since $-1 < A < +1$ is found everywhere except at
these spikes.
(The terms neglected in~(\ref{eq:approxinv}a) turn out to be
negative at the spikes.)
The behaviour of~$\invb$ is similar, except for the influence of
$D$ in equation~(\ref{eq:approxinv}b).
At each positive spike in~$P$ there is a zero of~$D$ together with
an odd number of sign changes in~$\invb$.
Zeros of~$D$ at points where no spikes develop produce
isolated sign changes in~$\invb$.
However, although a pair of zeros appears in~$D$ where a double spike
forms in~$Q$, no unusual behaviour occurs in~$\invb$ there.
It seems that zeros in~$A$ counteract the effect of the zeros in~$D$
sufficiently so that terms neglected in equation~(\ref{eq:approxinv}b)
become significant and prevent~$\invb$ from changing sign.

(We note that the sign change clusters in~$\inva$ and~$\invb$
evolve at different rates, and this leads to some discrepancies
between figure~4 and the behaviour described in the paragraph 
above.
By time $\tau = 6\pi$ some of the clusters are already too narrow
to be resolved and inaccuracies at the double spikes are large
enough to cause the invariants to behave erratically there.)

It follows that the formation of positive spikes in~$P$ is more than
a coordinate effect
induced by the form of the Gowdy metric, and is connected with unusual
behaviour of the spacetime curvature.
On the other hand our results suggest that this may not be the case 
for the double spikes that form in~$Q$:
while both~$P$ and~$Q$ can be related to the proper lengths of the
trajectories of the metric's~$T^2$ isometry group, the spiky
behaviour of~$Q$ does not seem to have a physical effect beyond this.

If we extrapolate the observed behaviour of the curvature
invariants forward in time to the singularity at $\tau = \infty$
we see that they become infinite---except possibly at
isolated values of~$\theta$ corresponding to the positive spikes
in~$P$.
This is in agreement with the curvature behaviour at the
singularity conjectured by Moncrief~\cite{M81}.
(The same reference shows that no curvature singularities
can develop before $\tau = \infty$,  
and our results do not suggest otherwise.)

\bigskip

The main features of the behaviour reported above are also
found in simulations using initial data other than
equation~\eqref{eq:cauchy} or a starting time other than
$\tau = 0$, and the implication is that the behaviour is
reasonably generic.
It is interesting to ask, though, what behaviour the spacetime
must to have prior to the starting time in order that it
arrives at the initial state we have chosen---that is,
what happens if we start with initial data~\eqref{eq:cauchy}
but reverse the direction of time?
We find that, in general, the variables~$P$ and~$Q$ have a
simple wave motion away from the singularity, as
equations~\eqref{eq:evol} suggest, but that for a starting time
closer to the singularity (e.g.,~$\tau > 4$) or a large
value for the parameter~$v_0$ in~\eqref{eq:cauchy} small-scale
spatial structure again starts to form.
(This is illustrated in figure~5.)
However, in this case the
small-scale structure does not grow indefinitely and
eventually the wave motion takes over.
Even when small-scale structure is developing, the curvature
invariants are found to be rapidly approaching zero,
and the spacetime `flattens' as it expands.

\begin{figure}[tb]
  \begin{center}
    \vskip 4mm
    \vbox {
      \beginlabels
      \refpos 90 725 {}
      \put 138 726 {\tau = -0.13}      
      \put 241 726 {\tau = 1.25}      
      \put 340 726 {\tau = 1.75}      
      \put 439 726 {\tau = 3.38}      
      \put 95 683 {P}      
      \put 159 634 {\theta}
      \put 258 634 {\theta}
      \put 356 634 {\theta}
      \put 455 634 {\theta}
      \endlabels
      \includegraphics[width=14cm]{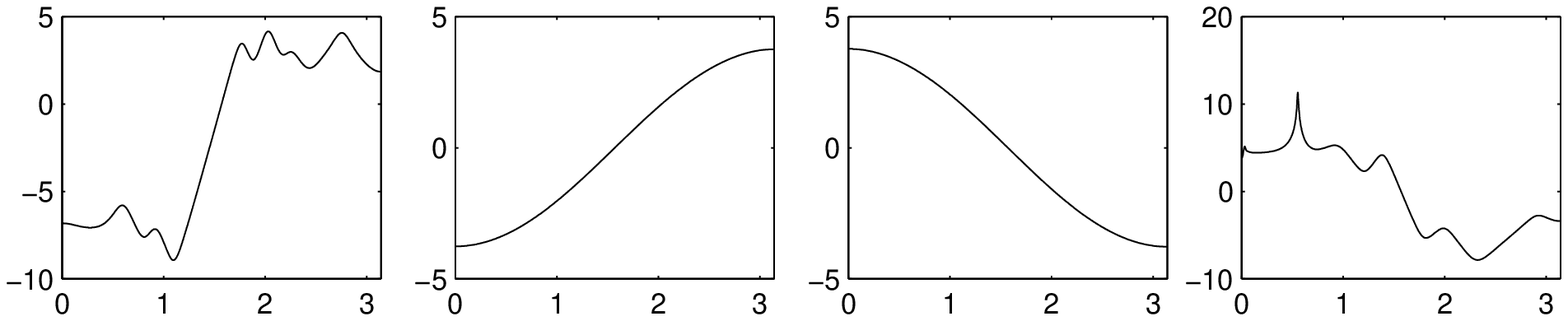}
    }
    \vskip 3mm
    \begin{quote}
      \small Figure 5:
      Development of small-scale structure in~$P$ for evolution both
      forwards and backwards in time.
      Initial data~\eqref{eq:cauchy} with $v_0 = 15$ is used at
      starting time $\tau = 1.5$.
      The left two plots are produced by evolving backwards from this
      initial state, the right two plots by evolving forwards.
    \end{quote}
  \end{center}
\end{figure}

\section{The AVTD and VTDB Hypotheses}
\label{sec:5}

Although the VTDB hypothesis of Eardley \emph{et al}~\cite{ELS} is
difficult to formulate in a covariant manner, it is easy to interpret
it for this problem.
In our evolution equations~\eqref{eq:con2} and~\eqref{eq:evol}
(combined in~\eqref{eq:first})
every spatial derivative is multiplied by a factor $\exp(-\tau)$.
Since the singularity is given by $\tau \rightarrow +\infty$,
it might seem reasonable to ignore these terms when studying the
singularity.
The resulting system of ordinary differential equations can be
solved analytically and one finds
\begin{equation}
  \label{eq:asymp}
  P(\tau,\theta) \sim a_0(\theta)\tau
  , \quad
  Q(\tau,\theta) \sim q_0(\theta)
  \qquad
  \hbox{as }\tau \rightarrow \infty.
\end{equation}
The numerical evidence in favour of this asymptotic behaviour is
compelling: $B \equiv Q_\tau$ and $A_\tau \equiv P_{\tau\tau}$
tend to zero as~$\tau$ increases.
Thus we can confirm Berger and Moncrief's assertion in~\cite{BM}
that the Gowdy~$T^3$ cosmologies exhibit AVTD behaviour.

\begin{figure}[htb]
  \hfil
  \hbox to 5.7cm {
    \vbox{
      \beginlabels
      \refpos 135 734 {}
      \put 126 626 {\tau}      
      \put 208 509 {\theta}
      \endlabels
      \includegraphics[width=5cm]{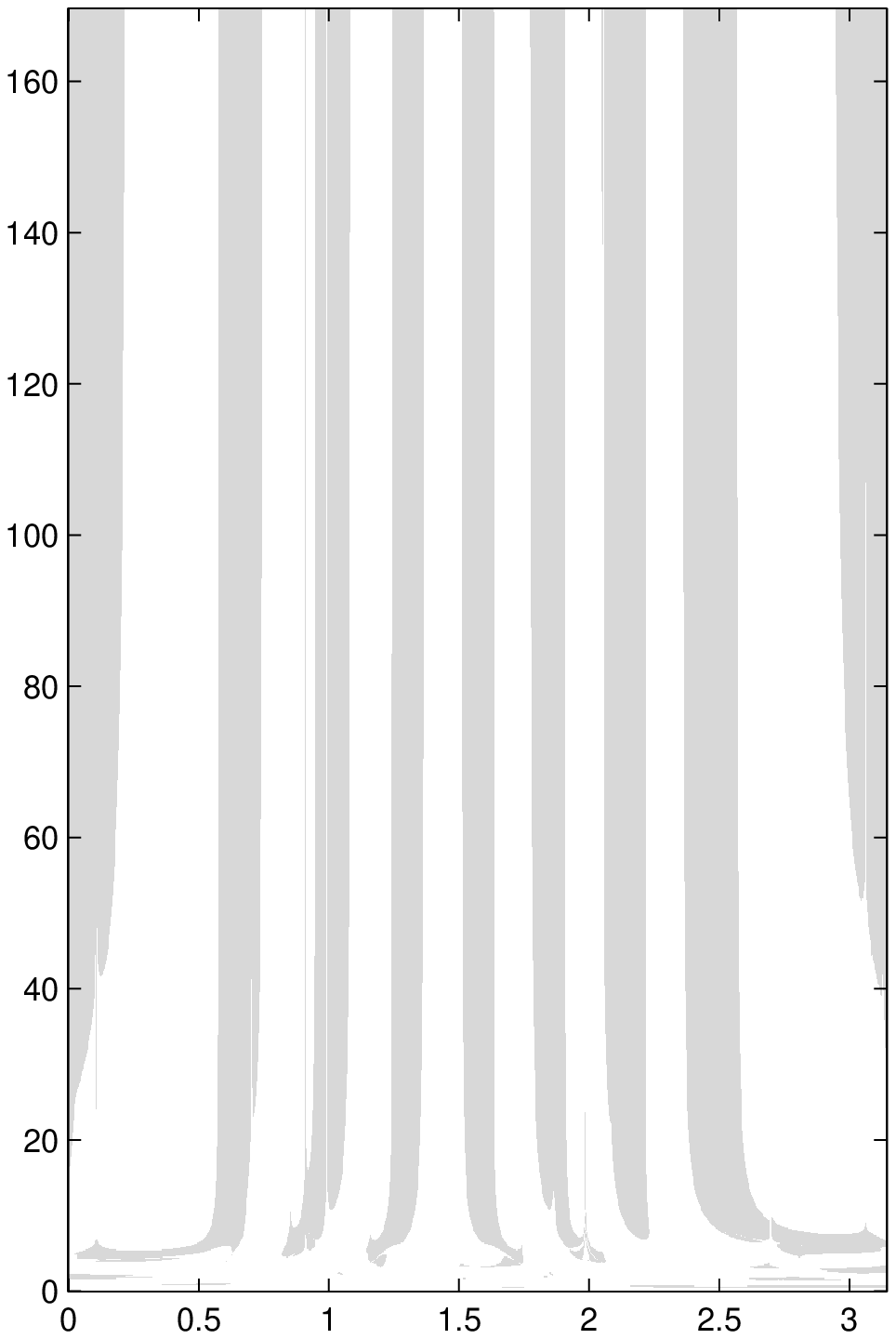}
    }
    \hss
  }
  \hbox to 7cm {
    \vbox{
      \vss
      \begin{minipage}[t]{7cm}
        \small Figure 6: 
        The VTDB region for $\tau \leq 54 \pi$.
        Spatial derivatives are considered negligible in the
        shaded regions, which correspond to the intersection
        of the light grey regions from both halves of
        figure~3.
      \end{minipage}
      \vskip 0.3cm
    }
    \hss
  }
  \hfil
\end{figure}

However, there is no evidence that the VTDB hypothesis used to 
derive~\eqref{eq:asymp} is valid!
The light-grey regions in figure~3 show where spatial derivative
terms are unimportant in the two wave equations, and
examination of their intersection reveals that 
for $\tau \leq 6 \pi$ there are
intervals of~$\theta$ in which VTDB occurs,
but that they do not cover the whole range of~$\theta$.
Figure~6 shows these VTDB regions for an extended time range, 
$\tau \leq 54 \pi$.
Clearly the behaviour is not universal.

In a recent preprint, Kichenassamy and Rendall~\cite{KR} have examined
analytically the behaviour of the Gowdy~$T^3$ cosmologies near the
singularity.
Their results seem to suggest that the asymptotic
behaviour~\eqref{eq:asymp} is universal, and is therefore not
contingent on the VTDB hypothesis.


\section{Conclusions}
\label{sec:6}

We have two groups of conclusions, one for cosmology and
one for numerical relativity.
We have investigated structure formation and behaviour near the
singularity in the Gowdy~$T^3$ cosmologies.
\begin{itemize}
\item  Our calculations are consistent with the earlier ones
       of Berger and Moncrief~\cite{BM}, but show more 
       fine-scale structure.
\item  This fine-scale structure is not a coordinate effect
       since it also occurs in the curvature invariants.
\item  As claimed earlier there is considerable numerical
       evidence that the AVTD conjecture is valid for the
       Gowdy~$T^3$ cosmologies, but there is no evidence to
       support the hypothesis of VTDB.
       Although spatial derivative terms matter, this does not
       seem to affect the asymptotic evolution near the 
       singularity.
\item  The Gowdy~$T^3$ cosmologies have the unexpected property
       that smooth initial data acquires structure when it is
       evolved both backwards and forwards in time.
\end{itemize}

We have not investigated in detail the symplectic integrator
used by Berger and Moncrief, but for this problem it would
seem to perform no better than standard finite-difference methods.
A simple Lax-Wendroff code using 8000 grid points without AMR will
produce results that differ from those shown in figure~1 only
at the points where spiky features develop.
However, the AMR code is faster and allows details smaller
than the coarse grid spacing to be investigated.
Summarizing the advantages AMR brings for this problem:
\begin{itemize}
\item  Results (on the coarsest grid) with the required level of 
       overall accuracy are produced in minimal time.
\item  Small spatial features are resolved without requiring
       a high density of grid points everywhere on the domain.
\item  The total number of grid points changes with time
       and allows quick evolution of the `frozen' spatial
       structures found at late times.
\item  No prediction as to the amount of fine-scale structure
       that will develop is needed beforehand.
\end{itemize}

There is another, subtle feature of AMR.
We have been asked frequently why Lax-Wendroff with AMR gives 
results which are indistinguishable from the highly
sophisticated \textsc{clawpack} codes.
At first sight the Lax-Wendroff scheme would appear to be 
inappropriate:
it contains artificial viscosity which would smear out the spatial
structure we observe.
However the viscosity coefficient is~$O((\Delta \theta)^2)$ 
where~$\Delta \theta$ is the spatial grid spacing, and
when spatial structure is encountered the AMR code evolves it with
a very small~$\Delta \theta$ so that consequently there is very 
little dissipation.

\bigskip

We thank Neil Cornish for useful discussions,
and Beverly Berger, David Garfinkle and Vincent Moncrief
for encouragement.
Simon Hern was supported by an EPSRC studentship.

\bibliographystyle{plain}
\end{document}